\documentclass[aps,prl,amsmath,amssymb,reprint,superscriptaddress,twocolumn,showpacs]{revtex4-1}

 \usepackage{graphicx}
 \usepackage{amsmath}
 \usepackage{amsfonts}
 \usepackage{amssymb}
 \usepackage{pstricks}
 \usepackage{psfrag}
 \usepackage{epsfig}
 \usepackage{changes}
 \usepackage[ansinew]{inputenc}
\usepackage{mathptmx}
\usepackage{helvet}
\usepackage{textcomp}
\usepackage{ulem}

\begin{document}


\title{Blood crystal:  emergent order of red blood cells under wall-confined shear flow}



\author{Zaiyi Shen}
\affiliation{Univ. Grenoble Alpes, LIPHY, F-38000 Grenoble, France}
\affiliation{CNRS, LIPHY, F-38000 Grenoble, France}
\author{Thomas M. Fischer}
\affiliation{Univ. Grenoble Alpes, LIPHY, F-38000 Grenoble, France}
\affiliation{CNRS, LIPHY, F-38000 Grenoble, France}
\affiliation{Laboratory for Red Cell Rheology, 52134 Herzogenrath, Germany}
\author{Alexander Farutin}
\affiliation{Univ. Grenoble Alpes, LIPHY, F-38000 Grenoble, France}
\affiliation{CNRS, LIPHY, F-38000 Grenoble, France}
\author{Petia M. Vlahovska}
\affiliation{Engineering Sciences and Applied Math, Northwestern University, Evanston 60208, USA}
\author{Jens Harting}
\affiliation{Helmholtz Institute Erlangen-N\"urnberg for Renewable Energy (IEK-11), Forschungszentrum J\"ulich, F\"urther Strasse 248, 90429 N\"urnberg, Germany}
\affiliation{Department of Applied Physics, Eindhoven University of Technology, P.O. Box 513, 5600MB Eindhoven, The Netherlands}
\author{Chaouqi Misbah}
\email[]{chaouqi.misbah@univ-grenoble-alpes.fr}
\affiliation{Univ. Grenoble Alpes, LIPHY, F-38000 Grenoble, France}
\affiliation{CNRS, LIPHY, F-38000 Grenoble, France}


\date{\today}

\begin{abstract}
Driven or active suspensions can display fascinating collective behavior, where  coherent motions or structures arise on a scale much larger than that of the constituent particles.
Here, we report experiments and numerical simulations revealing  that red blood cells (RBCs) assemble into regular patterns in a confined shear flow. The order is of pure hydrodynamic and inertialess origin, and emerges from a subtle interplay between (i)   hydrodynamic repulsion by the bounding walls which drives deformable cells towards the channel mid-plane and (ii)  intercellular hydrodynamic interactions which can be attractive or repulsive  depending on cell-cell separation. Various crystal-like structures arise depending on RBC concentration and confinement. Hardened RBCs in experiments and rigid particles in simulations  remain disordered under the same conditions where deformable RBCs form regular patterns, highlighting the intimate link between particle deformability and the emergence of order. The difference  in structuring ability of healthy (deformable) and diseased (stiff) RBCs creates a flow signature  potentially exploitable  for diagnosis of blood pathologies.

\end{abstract}


\maketitle


\paragraph{Introduction.---}
Blood is  a dense suspension of red blood cells (RBCs) of about 45\% volume fraction.  Blood  microstructure (RBC shapes and spatio-temporal distribution) strongly depends on flow conditions (local shear, flow curvature, vessel-cell size ratio) \cite{Popel:2005, Suresh2006, AbkarianBMrev:2008, Abkarian:2008, VlahovskaCR, Fedosov:2011, Li-Vlahovska:2012, Dupire:2012}. For example, in capillary flows  the RBC can adopt variety of morphologies, distinctly different than its biconcave equilibrium shape \cite{Lanotte:2016}. Parachute-shaped RBCs can arrange in a single  file of regularly spaced cells
\cite{McWhirter-Hiroshi-Gompper:2009, McWhirter:2012, Tomaiuolo:2012},  a pattern also observed with droplets in microfluidic inertialess flows \cite{beatus2006phonons, beatus2007anomalous,beatus2012physics, uspal2012collective, janssen2012collective,shani2014long,uspal2014self}.  Confinement is one factor promoting the order in these systems. If the bounding walls are removed then the  particles can pass over each other and  experience
  hydrodynamical diffusion (a cross-flow displacement after collision)  \cite{Cunha1996,zhao2012,grandchamp2013,rivera2015} that, very much like  classical diffusion, favors homogenization (intermixing of suspended
  entities) and tends to destroy any order.  Another factor that  gives rise to complex collective dynamics and order is  the long-range hydrodynamic interactions between  the fluid-embedded particles.  These interactions, i.e., correlations in the motions of particles mediated
by flows in the suspending liquid, underly many self-organizing phenomena, from  the microfluidic crystals \cite{beatus2007anomalous, baron2008hydrodynamic, lee2010dynamic, humphry2010axial, Shani:2014}  to directed macroscopic motion in self-rotating particles \cite{Bartolo:2013, Bartolo:2015, Yeo:2015, goto2015purely}.

Unlike for pressure-driven flows, structuring of  RBCs (or drops and capsules)  under confined shear flows has not been studied even though there are experiments showing formation of trains in RBC suspensions
\cite{RBCtrain, Bull:1983} and emulsions \cite{pathak2002layered}.
To fill this void,   we carried out  experiments and    3D numerical simulations to study the formation of  two-dimensional arrays (crystals)  by RBCs sheared   between two parallel plates.
 We observe ordering of RBCs.  Intriguingly the order persists even when the size of the gap allows cells to pass over each other, i.e., the emergence of order does not require strong confinement.
  We develop an analytical model that provides a physical insight into the phenomena.
  The  pattern wavelength (cell-cell separation) predicted by the model is in excellent agreement  with simulations and experiments.

\begin{figure}
\centering
\includegraphics[width=0.8\columnwidth]{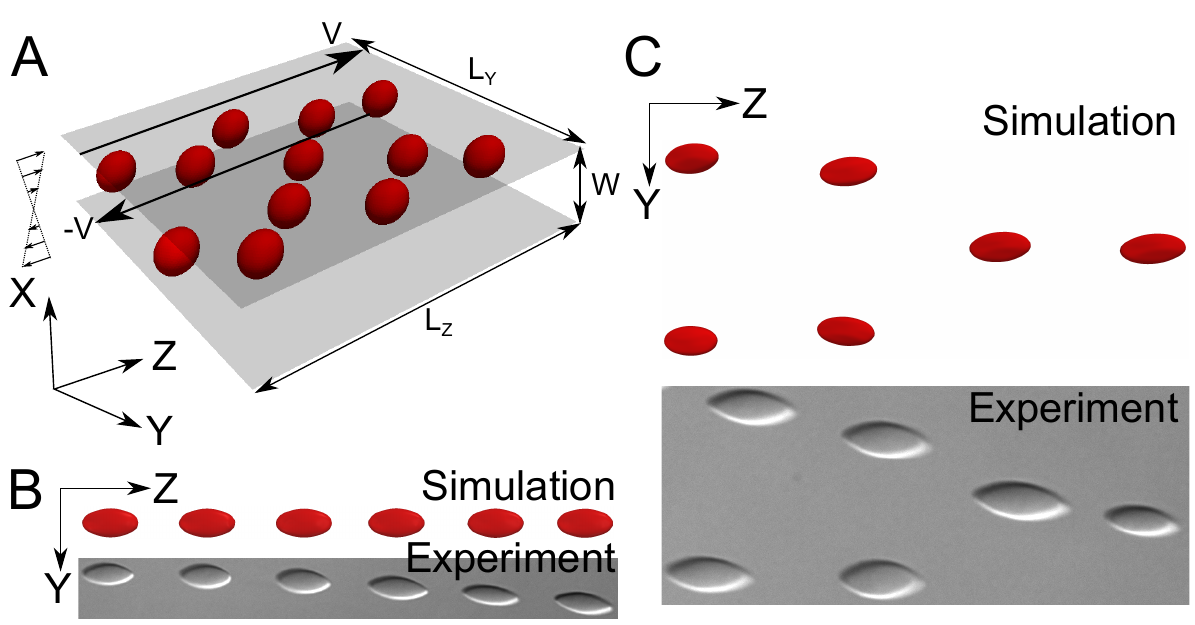}
\caption{ \footnotesize(A) (color online) A schematic view of the  simulation set-up.
(B) Both simulation and experiment show RBCs organized in stable chains along the flow direction; $W=3R$.
The computational domain is $L_Y=12R, L_Z=30R$. {(C) Chains can merge, resembling crystal dislocation.} \label{fig1}}
\end{figure}
\paragraph{The model.---}
The lattice Boltzmann method (LBM) \cite{Krueger2013, kruger2014} was used to solve the quasi-incompressible Navier-Stokes equations for fluid flow.
\textcolor{black}{ The membrane energy  consists of a contribution due to resistance to bending $(\kappa/2) H^2 $, with $H$ the mean curvature and $\kappa$ the bending rigidity modulus, and resistance to in-plane shearing and stretching $\kappa_s(I_1^2 + 2I_1-I_2)/12+  \kappa_\alpha I_2^2/12$,
where $\kappa_s$ is the shear elastic modulus, $\kappa_\alpha$ is the area dilation modulus, and
$I_1$ and $I_2$ are the in-plane strain invariants (see \cite{Krueger2013}).
$\kappa_\alpha/\kappa_s=200$ is chosen large enough to preserve local area conservation.}
The RBC size $R$ is taken to be  $R\simeq 4\mu$m.
The reduced volume is defined as $\nu=(3V/4\pi) (A/4\pi) ^{-3/2}$, with $A$ the particle surface area and $V$ its volume. $\nu=0.64$ was used for RBC model when compared with experiments, otherwise $\nu=0.98$ was used (in this case we refer to the cell as a capsule model). This allowed computational efficiency without affecting the main results (see below). $C_{as}=\eta_0\dot\gamma R/\kappa_s$ is the capillary number. 
$C_{as}$ is set to 0.05. 
 The viscosity contrast is fixed to one. The reference shape is taken to be  to the equilibrium one in the absence of shear elasticity.
The suspension is sheared between two parallel planes at a constant shear rate $\dot\gamma=2V/W$, where $2V$ is the relative velocity of the planes and $W$ the channel width (Fig.\,\ref{fig1}A).  

\paragraph{The experimental set-up.---}
The experiments were performed in a home-made rheoscope with cone-plate geometry.
Microscopic images were taken with a CCD camera (DMK 41BF02.H, The Imaging Source Europe GmbH, Bremen, Germany).
Normal blood samples were obtained from the EFS (Etablissement Fran\c{c}ais du Sang) and kept refrigerated until use. Solid spherical particles were produced by suspending RBCs in an isotonic solution of sodium salicylate (Sigma-Aldrich) thus converting the biconcave RBCs into spheroechinocytes. This shape was then conserved by fixation with 0.25\% glutaraldehyde (Alfa Aesar, Karlsruhe, Germany). RBCs and rigid spheres were washed three times with isotonic PBS (Dulbecco, biowest, Nuaille, France). RBCs or spheres were suspended in an isotonic solution of dextran (MW 500000 D, Sigma-Aldrich, Saint-Quentin Fallavier, France) plus PBS. The viscosity at room temperature (25$^{\circ}$C) was 50mPas (Anton Paar, Rheoplus, Graz, Austria), and for some experiments (as shown in \cite{SM}), unless otherwise indicated. Suspensions of RBCs or spheres were prepared with volume fractions between 0.002 and 0.01 and were loaded into the cone-plate chamber.  The shear rate was varied from 15 to 94 $s^{-1}$. The viscosity of the hemoglobin solution of healthy RBCs is around 10 mPa.s at room temperature (25$^{\circ}$C). The experimental capillary number is in the range $C_{as}\simeq 0.75-3$.

\paragraph{ Flow-aligned chains.}---
RBCs in shear flow display two main types of dynamics depending on the applied shear stress:  Tank-treading (TT) at large shear stress, where the cell assumes stable orientation relative to applied shear direction,  and   tumbling at low enough shear stress, where the cell executes periodic flipping motion \cite{Fischer894}.
In both simulations and experiments the parameters are chosen  such that cells are in the TT regime. Experiments and simulations show the formation of regularly spaced chains of RBCs aligned with the flow direction, see  Fig.\,\ref{fig1}B and {\textcolor{black}{Movie 1 and 2 in \cite{SM}}}.

Starting from a random initial cell distribution, we observe a transient regime during which cells mix (hydrodynamical diffusion) due to cell-cell hydrodynamic interactions and migrate towards the midplane due to the hydrodynamic repulsion by the wall. Once all cells have reached the channel midplane, the degree of disorder decreases continuously until the cells reach an ultimate stable configuration of ordered
chains (Fig.\,\ref{fig1}B).  Both experiments and simulations show that chains can merge into a stable Y-configuration (Fig.\,\ref{fig1}C).

The finite ratio between channel width $W$ and  cell size $R$ is a crucial factor in the cell structuring. We  find that stable order is impossible if the cells were considered as points.
An intriguing  observation (analyzed in detail below) is that order persists even for weakly confined suspensions (with the gap between planes about ten times the cell radius $R$), where one would have  expected that
cell-cell hydrodynamic interactions (responsible for hydrodynamic diffusion)  pushing the cells out of the channel center  would allow the imposed shear flow to advect the cells further apart, thereby favoring disorder.

\begin{figure}
\centering
\includegraphics[width=0.8\columnwidth]{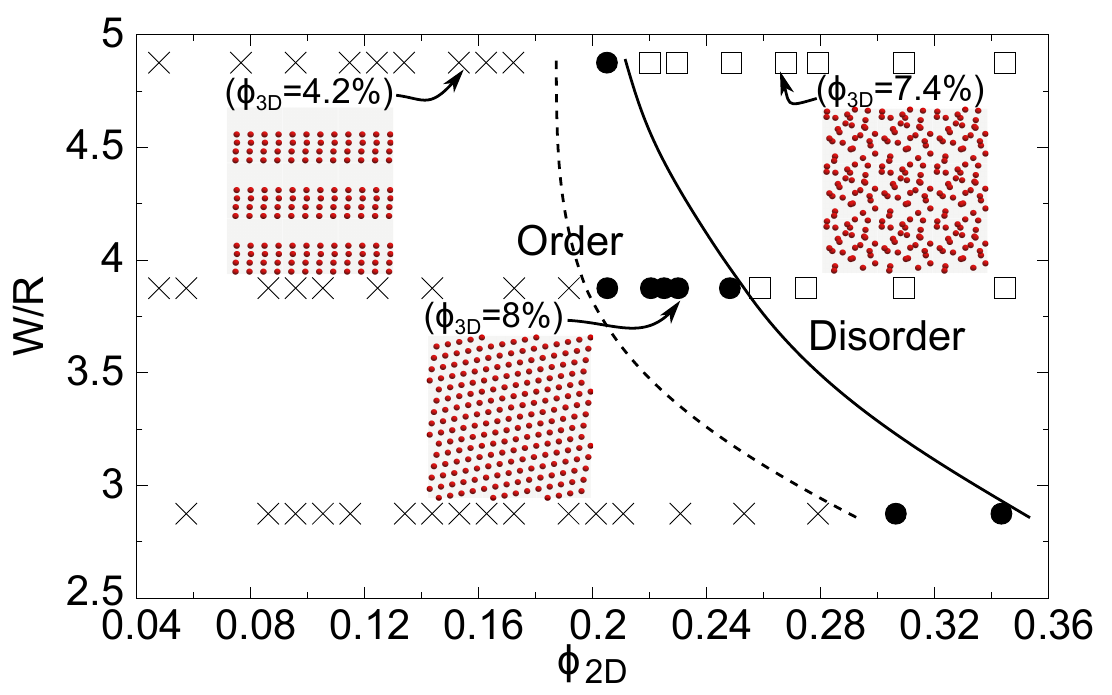}
\caption{ \footnotesize (color online) Phase diagram of order and disorder varying RBC concentration and confinement.
  $\phi_{2D} =   N R^2 \pi / (L_Y L_Z)$, where $N$ is the number of particles.
  The computational domain in $Y$ and $Z$ direction is varied from $8R$ to $18R$ for different cases.
  The computational domain for the insets is $L_Y=L_Z=18R$.
  Displayed are 3x3 computational domains.
  \label{fig4}}
\end{figure}

\paragraph{ Crystals.---}
 Numerical  simulations  show that at low  particle volume fraction,  capsules form flow-aligned chains. The spacing between the chains in the vorticity direction also shows periodicity.  The chains slide relative to each other due to a slight displacement in $X$ (velocity gradient)-direction and advection by the imposed shear flow (see Movie 3 in \cite{SM}). Interestingly, the chain offset along $X$ fluctuates in time but despite these fluctuations the order in the $Z$ (flow) direction  persists (see also Fig. 1 in \cite{SM}).
The chaining  occurs when starting from a random RBC/capsule distribution in 3D. Two closely  located  parallel chains  repel each other but still preserve their structure  (see Movie 4 in \cite{SM}). Increasing the volume fraction  results in formation of infinite 2D lattices. The crystal configuration assembles from random initial conditions (Movie 5 in \cite{SM}). At  high volume fraction disorder prevails. Fig.\,\ref{fig4} summarizes the  phase behavior of the 2D crystals.

\begin{figure}
\centering
\includegraphics[width=0.8\columnwidth]{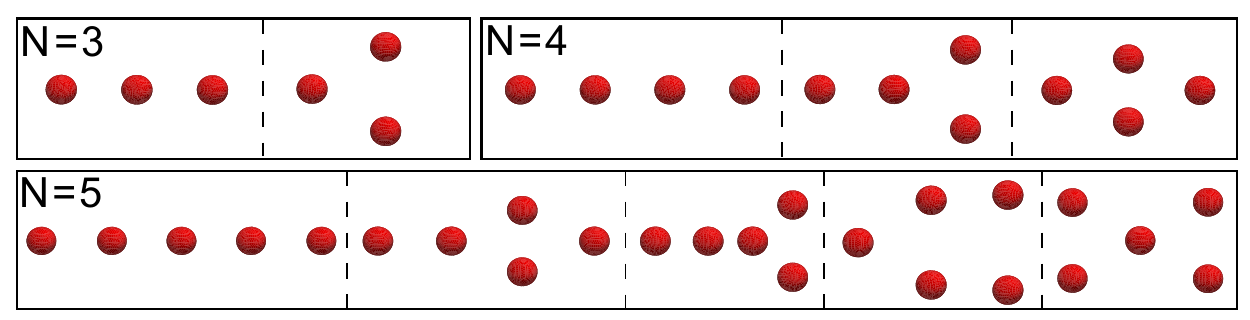}
\caption{ \footnotesize (color online)
Ordered stable patterns (in $ZY$ projection) of 3, 4 and 5 interacting particles with confinement of $W=2.9R$. The computational domain  is $L_Y=18R, L_Z=25R$. \label{figYshape}}
\end{figure}

In addition to linear chains we discovered numerically other  configurations.
We explored the stability of other possible  configurations  of three, four and five capsules, as shown in Fig.\,\ref{figYshape}, in order to probe the existence of crystalline  structures other   than 1D chains.
 {The simulations show that the final stable configurations are all symmetric about the flow direction. }
These elementary crystal configurations serve as building blocks to larger crystals and imply the existence of two types of crystals (a) a 1D crystal corresponding to an infinite chain and (b) a 2D crystal based on the triangular arrangements in Fig.\,\ref{figYshape}. This triangular arrangement is also observed in experiments (Fig.\,1C and Fig.\,1 in \cite{SM} for the whole image).

\begin{figure}
\centering
\includegraphics[width=0.8\columnwidth]{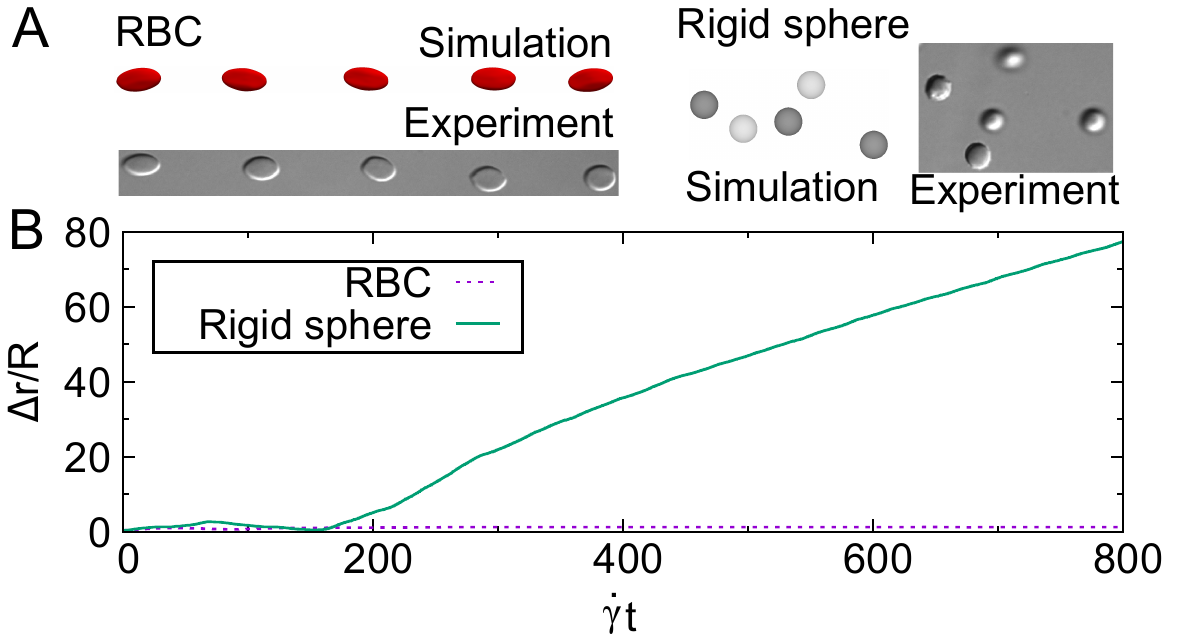}
\caption{ \footnotesize(A) (color online) Configurations for deformable cells after long time: Simulation, $W=4.9R$, $Re=0.05$, $L_Y=12R, L_Z=37.5R$; Experiment, $W=4.9R$, $C_{as}=0.7$, $Re \sim 10^{-5}$. Configurations for rigid spheres after long time: Simulation, $W=4.5R$, $Re=0.05$, $L_Y=L_Z=18R$; Experiment, $W=4.5R$, $Re \sim 10^{-5}$. (B)
Distance between two particles as a function of time: dashed line corresponds to deformable cells, while solid line corresponds to rigid spheres, or hardened RBCs.
\label{fig5}}
\end{figure}

\paragraph{Effect of cell deformability.---}
An important ingredient for the emergence of order to be discussed below is the wall-induced migration  which requires cell deformation  \cite{Cantat1999b,seifert1999,sukumaran2001}. For a spherical particle, there is no cross-streamline migration in the Stokes regime, owing to the linearity of the Stokes equations. We analyzed the impact of cell deformability on the emergence of order (Fig.\ref{fig5}).
Our simulations show that  rigid particles
never settle in the midplane (due to the absence of wall-induced migration), and disorder prevails; solid spheres even if initially placed on the midplane, drift apart. This is illustrated by the simulation in Fig.\ref{fig5}  where  disordered pattern is obtained.
  Moreover, Fig. \ref{fig5}B shows
that the distance between two given rigid particles increases  with time without saturation (solid line), whereas the same quantity shows saturation (indicating stable pairing) when deformable cells are considered (dashed line).

Experiments on hardened RBCs  also confirm the lack of order (Fig.\ref{fig5} A). These results  support the idea that the wall-induced migration {due to particle deformability}  plays a crucial role in the ordering process.  Note that in  the presence of inertia, even a spherical particle will undergo a wall-induced migration allowing   ordered patterns of rigid particles to be stabilized\cite{humphry2010axial}. Inertia is also responsible for  hydrodynamic ordering of rotating disks \cite{goto2015purely} and strong focalization of capsule suspensions \cite{kruger2014} in a pressure-driven flow. Inertia, however, is negligible in our study.

\begin{figure}
\centering
\includegraphics[width=0.8\columnwidth]{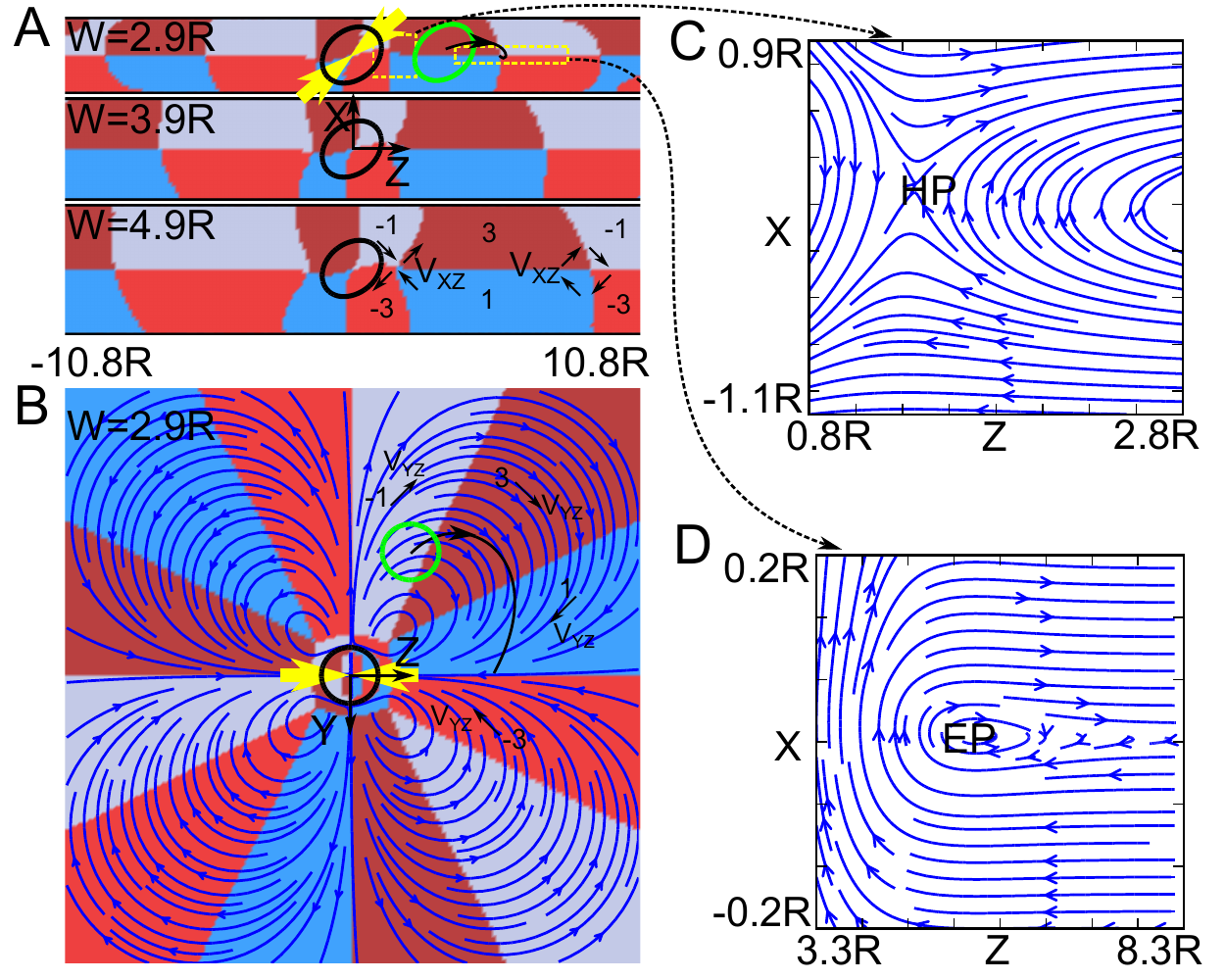}
\caption{ \footnotesize (color online) (A) In the shear plane, the color code shows the value of $Col=V_Z/\vert V_Z \vert + 2V_X/\vert V_X \vert$. The gray, dark red, blue and red colors correspond to $Col=-1, 3, 1, -3$ respectively.  $-1$ refers to $V_X<0$ and  $V_Z>0$, 3 to $V_X>0$ and $V_Z>0$, 1 to $V_X>0$ and  $V_Z<0$ and  -3 to $V_X<0$ and  $V_Z<0$.
The intersections of the four colors show alternatively EPs and HPs. The streamlines at EP and HP are shown in the right panel (C and D). The black solid line with arrow shows the schematic trajectory of the cell represented by the green contour, interacting with the cell represented by the black contour. The computational domain is $L_Y=36R, L_Z=36R$ in all these cases.
(B) The flow field around a single cell performing TT in the midplane. The cell resists stretching by exerting a force dipole (yellow arrows). The color code shows $Col$ with $X$ being substituted by $Y$. \label{fig2}}
\end{figure}

\paragraph{Flow structure around a single cell and around a  pair.---}
Let us now focus on the basic understanding of the crystal formation.
We first analyze the flow field around a single cell.
Fig. \ref{fig2}B shows that in the Y-Z-plane this flow field is quadrupolar in nature. 
Fig. \ref{fig2}A
 shows the same flow in the $X$-$Z$(shear)-plane. Fig.\,\ref{fig2}D zooms into  the flow
 and shows  the existence of recirculation zones, nonexistent in unconfined shear flow \cite{zurita2007swapping, mcwhirter2009, janssen2012collective}. Their centers  are designated as { elliptic points (EPs) hereafter}.
The center of a TT cell also constitutes an EP.
Between two EPs there is a point where the flow locally is hyperbolic (Fig.\,\ref{fig2}C). This point is referred to as hyperbolic point (HP). In Fig.\,\ref{fig2}A,  these points  are located  where four colors meet.
Movie 6 in \cite{SM} illustrates experimentally the existence of an HP close to a single TT RBC.
The EPs are essential for the formation of RBC chains as suggested earlier \cite{RBCtrain}.

\begin{figure}
\centering
\includegraphics[width=0.8\columnwidth]{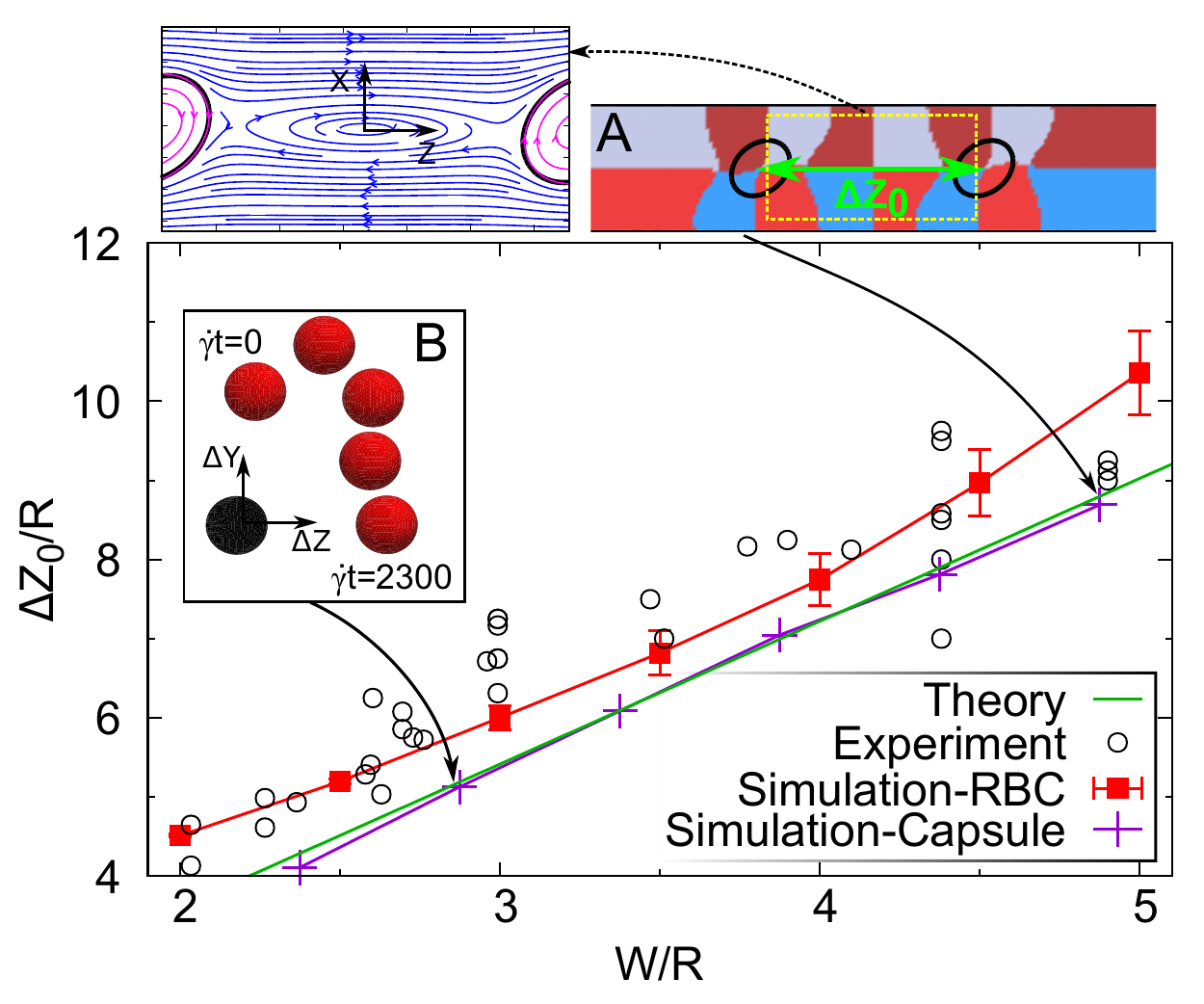}
\caption{  \footnotesize (color online) Equilibrium distance of two cells as a function of gap width $L_Y=18R, L_Z=36R$ for the simulations of capsules and $L_Y=12R, L_Z=27R$ for the simulations of RBCs. Error bars in simulation result from the fact that RBCs "swings" causing small oscillations of the equilibrium distance. Insets: (A) The flow field in shear plane when two capsules form a stable pair (capsules are shown by black ellipses). $L_Y=L_Z=36R$. And (B) the evolution of two capsules with time showing attraction and ultimate formation of a stable pair. Here $W=2.9R$, $L_Y=L_Z=18R$.
\label{fig3}}
\end{figure}

 Fig.\ref{fig3} shows an example of  the trajectory of pairing cells (B inset) and the flow field after a stable pair is formed (A left panel, see also Movie 7 in \cite{SM}).
Comparison with Fig.\,\ref{fig2}A shows (i) that in pair formation the second cell settles close to the EPs created by the first cell (see also analytical theory below) and (ii) that between a cell pair there are two HPs and one EP.

If pair formation is studied at different confinements an almost linear dependence of the equilibrium pair-distance $\Delta Z_0$
  with channel width is observed, Fig.\,\ref{fig3}. The experimental data is in good agreement with RBC simulations, and slightly above the capsule simulations and theory (see below). This indicates that the small reduced volume of RBCs causing their elongation   also contributes to the equilibrium distance.

\paragraph{ Analytical theory for pair formation.---}\label{sec:analytical}
Here we provide a theoretical explanation for the pair formation.
A detailed analysis of this  phenomenon (see theory in \cite{SM}) shows that the pairing results from an intricate interplay  between    (i) a long-range hydrodynamic attraction of two cells along the flow direction, (ii) the wall-induced  migration across the streamlines, (iii) a short range hydrodynamic repulsion between the cells due to the imposed shear flow (since the cell mass centers are not exactly on the midplane, their relative translational velocity is nonzero; thus the shear flow is acting to separate the pair).
As dictated by translational invariance along $Y$, the coordinates of the  cells in the pair can be written as $(-\Delta X,0,0)$ and $(\Delta X,0,\Delta Z).$ The question amounts to determining the steady state and stable positions $\Delta X_0$ and $\Delta Z_0$.

\begin{figure}
\centering
\includegraphics[width=0.8\columnwidth]{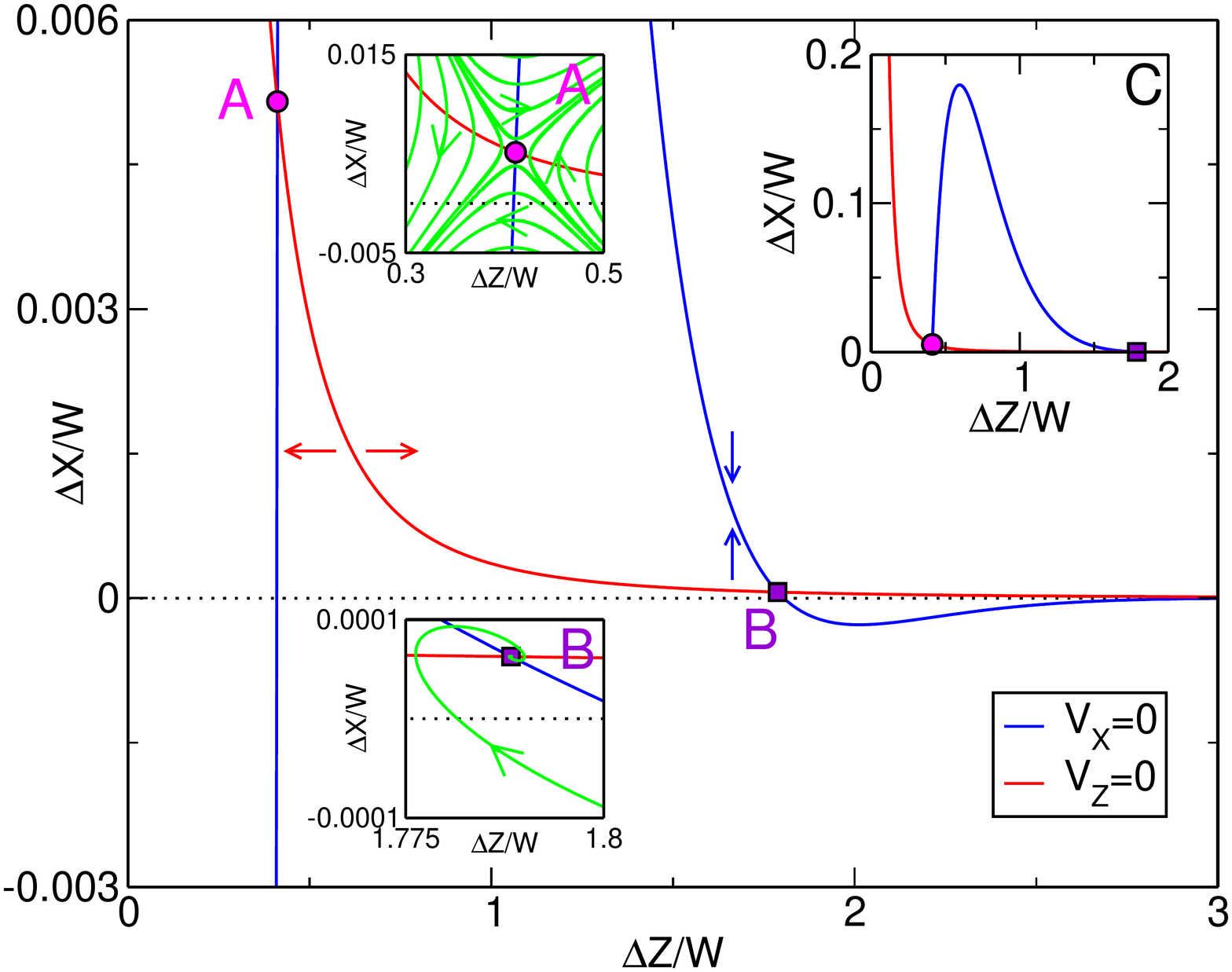}
\caption{  \footnotesize  (color online)
Theoretical analysis of relative motion of two cells in confined shear flow. The cells are located in positions $(-\Delta X,0,0)$ and $(\Delta X,0,\Delta Z).$ Blue (red) curve marks the separations for which the $X$- ($Z$-, respectively) component of the velocity of the cells vanishes. Crossing the blue (red) curve exchanges the cell interaction along the $X$- ($Z$-, respectively) direction between attraction and repulsion, as shown by the blue (red, respectively) arrows. The intersections of the red and the blue curves correspond to stationary separations. Magenta circle (point $A$) corresponds to the unstable separation. Violet square (point $B$) corresponds to the stable separation. Black dotted line corresponds to the case when both cells are in the midplane. Inset A: Orbits of the separation $(\Delta X(t),0,\Delta Z(t))$ in vicinity of the unstable stationary separation (point $A$). Arrows indicate the direction in which the separation evolves with time. Inset B: An orbit of the separation $(\Delta X(t),0,\Delta Z(t))$ in vicinity of the stable stationary separation (point $B$). Arrow shows that the separation converges to the stationary value with time. Inset C: Red and blue curves at a larger scale.
\label{fignew}}
\end{figure}

A steady-state solution corresponds
to the velocity of both cells being zero (the pair of cells is at rest in the laboratory frame).
The $Z$ (flow-direction)-component of the velocity of a cell in a pair has two contributions: (i) the velocity field induced by the first cell and (ii) the unperturbed shear flow.
The first effect can be well approximated by the quadrupolar flow field shown in Fig.\,\ref{fig2}B.
This flow field has a monotonic algebraic decay (see \cite{SM}) as the distance between the two cells increases.
The red curve in Fig.\,\ref{fignew} shows the location where the total (quadrupolar + imposed shear) contribution to the $Z$-component of the velocity vanishes in the  ($\Delta X$,  $\Delta Z$) plane.

The $X$ (velocity-gradient direction)-component of the velocity of a cell in the pair has two contributions: (i) the flow induced by the other cell and (ii) the wall-induced migration across the streamlines.
The first contribution has a complicated form \cite{Liron1976}, which can be well approximated by a rapidly decaying attenuated sine wave (Theory in \cite{SM}).
The second contribution is proportional to the cell displacement from the midplane  (Theory in \cite{SM}).
Equating the sum of these contributions to zero gives a second relation between $\Delta X$ and $\Delta Z,$ shown by  the blue curve in Fig.\,\ref{fignew}.

The intersections of the blue and the red curves in Fig.\,\ref{fignew} correspond to stationary separations of the two cells.
Two such intersections can be identified: The separation corresponding to the point $A$ in Fig.\,\ref{fignew} is unstable, as suggested by the orbits in Fig.\,\ref{fignew}A. The separation corresponding to the point $B$ is stable as suggested by the orbit in Fig.\,\ref{fignew}B.

For weak confinement $W/R\gg 1$  (channel gap large compared to the   cell size), we deduce an asymptotic scaling law  (see \cite{SM})
\begin{equation}\label{result1} \Delta Z_0/W=1.805\end{equation}
This scaling is universal as it is independent of the details of the cell structural parameters. Details of the physical nature of the cell (e.g. reduced volume, elastic properties, etc..), or its precise shape shows up  only in the correction terms \cite{SM} to the main scaling given above.
Equation (\ref{result1}) provides a very good agreement with the  full numerical simulation (Fig.\ref{fig3}).

We have analyzed the linear stability  of the stationary-pair solution and found  one real negative eigenvalue (by looking for perturbations in the form $e^{\omega t}$, with $\omega$ the eigenvalue), corresponding to the attraction along the $Y$ direction and two complex eigenvalues with negative real parts corresponding to the stability in the $X$-$Z$-plane, shown schematically in Fig.\,\ref{fignew}B. The complex nature of the eigenvalues leads to spiraling of the trajectory towards to the stable fixed point (Fig.\,\ref{fignew}B).

\paragraph{Conclusions and outlook.---}
{\color{black}{
We have studied the structures formed by RBC and capsule suspensions in weakly confined shear flow using experiments and numerical simulations with LBM.
Most of the presented simulations are performed for  quasi-spherical  capsules (for numerical efficiency). However, we have checked that the main features are captured for a reduced volume equal to that of a RBC (\textcolor{black}{see Movies 2 and 3 in \cite{SM}}).
Our work highlights the key roles of cell deformability and shape in the emergence of order. The analytical theory points to  cross-stream  migration of the cells as the driving force of cell-cell pairing and eventual multi-cell ordering.
Our study suggests that the structures formed by RBCs with different pathologies could be different; this could serve as a diagnostic tool since several blood diseases are accompanied by  decreased deformability  (sickle cell anemia) or shape change (elliptocytosis). To explore this idea,  more detailed, pathology-specific models of the RBC would be needed for the simulations.
 Cell-cell separation in a stable pair is found to be  linearly dependent on the gap width and, for nearly spherical capsules,   insensitive to the interfacial mechanics. This suggests that this result should apply to any other system  of soft particles experiencing a wall-induced migration, such as drops and vesicles.
}}

\pagebreak

\section{Supplementary Material}

\begin{figure}[h]
\centering
\includegraphics[width=1\columnwidth]{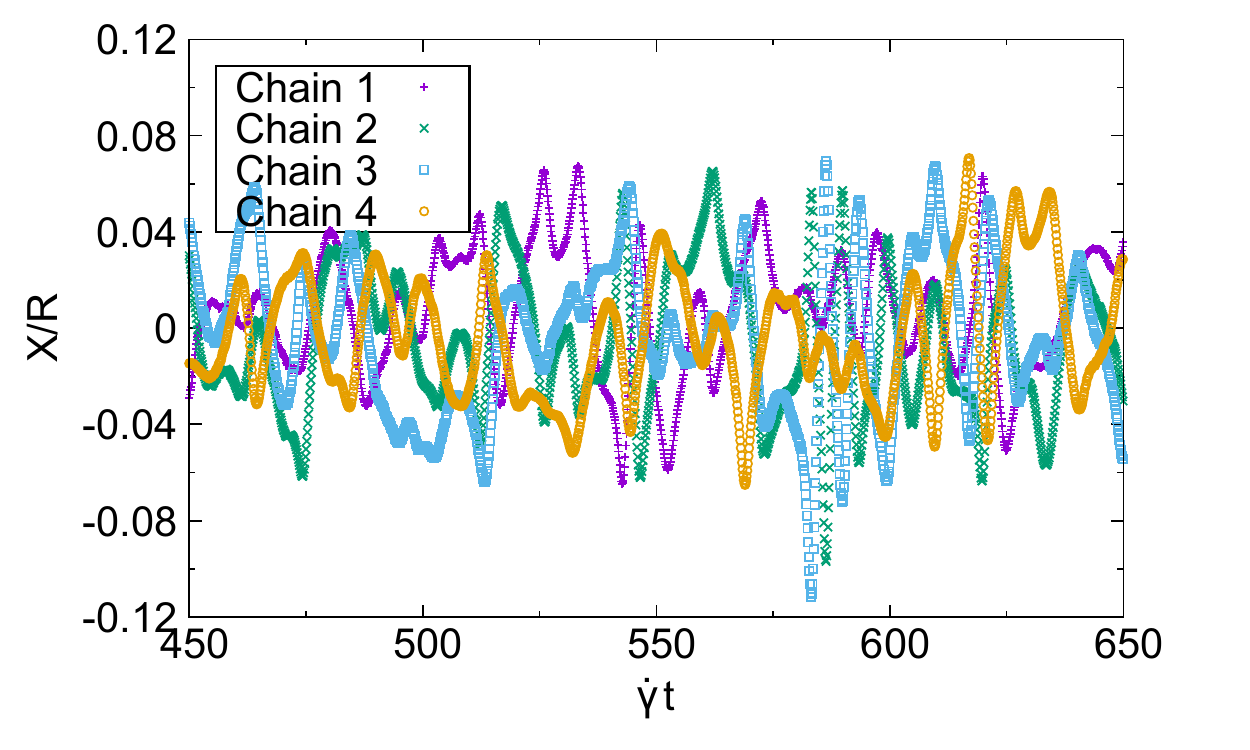}
\caption{\footnotesize The X position of each chain when four chains form. the positions fluctuate.}
\label{fig:chain_x}
\end{figure}

Here we show that particle positions fluctuate (FIG. \ref{fig:chain_x}), yet the crystal is stable (See Movie 2) in simulations.

\begin{figure}[tbhp!]
\centering
\includegraphics[width=0.8\columnwidth]{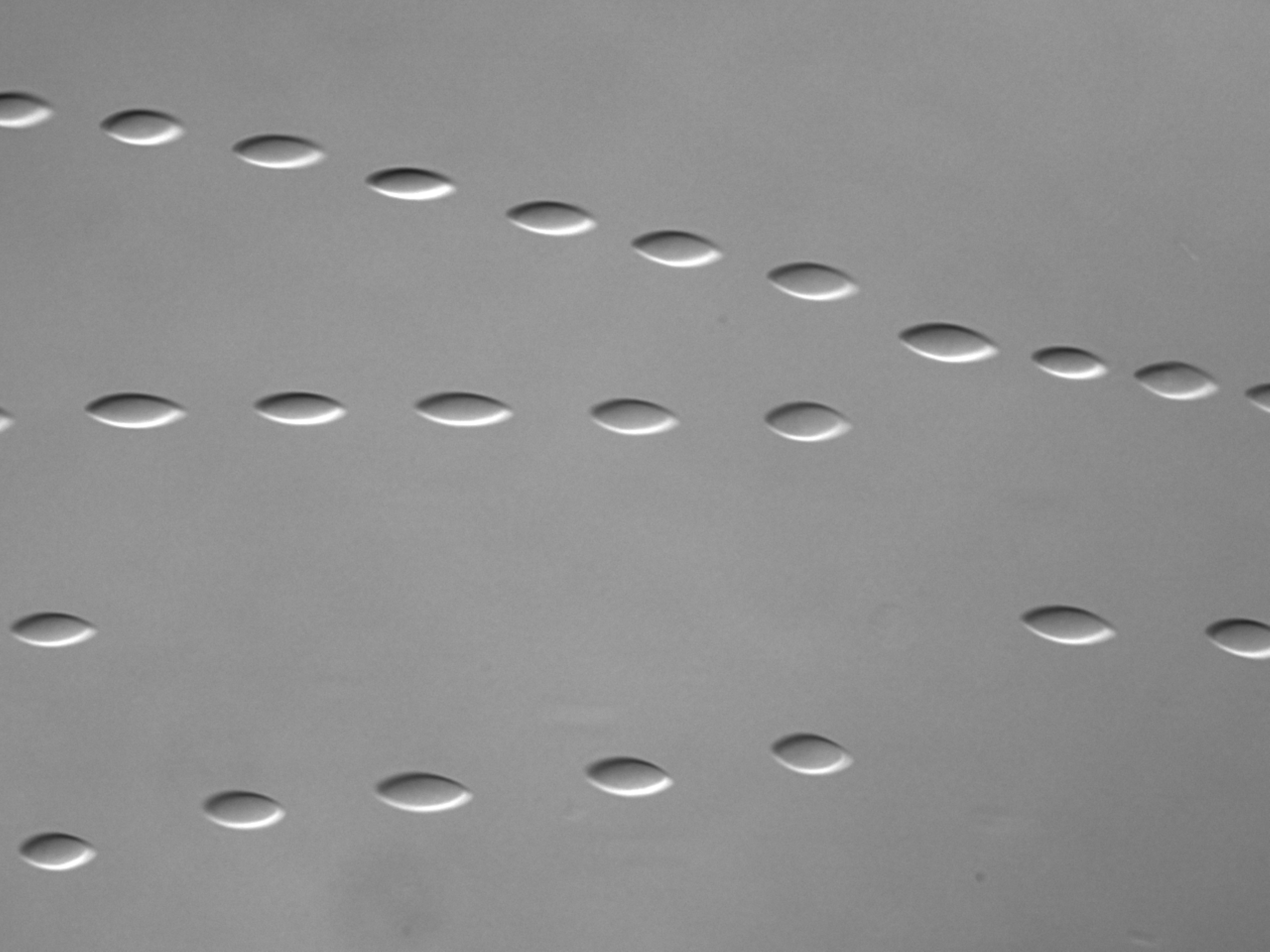}
\caption{\footnotesize Bifurcation of a chain, $\dot\gamma=$56/s, gap thickness=17$\mu$m.}
\label{fig:bifurcation}
\end{figure}

In experiment, stable Y-bifurcations form (FIG.\,\ref{fig:bifurcation}).

\subsection{Theory}

We start with an analytical model of the quadrupolar flow field of a single cell.
Along the flow direction, this field provides an attractive tendency for a second cell.
To explain the establishment of an equilibrium distance of the two cells, we invoke two additional mechanisms:
(i) the displacement of the second cell from the midplane due to the short-range flow disturbance by the first cell and (ii) the wall-induced migration due to the finite size of both cells.

Consider a single cell in a shear flow placed in the midplane at $X=Y=Z=0$. The cell
  has a fixed orientation, while its membrane undergoes a TT motion. Shear flow {contains}  a straining component that stretches the cell (see \cite{Petia_review}). The cell reacts by opposing this stretching (Fig.\,6 of the main text, yellow arrows). This effect is characterized by a stresslet, which is a rank-2 tensor. For analytical simplicity we represent it by a force dipole $\Sigma$. As will be seen later, this difference results only in different numerical prefactors of order unity, which do not affect the final answer.

From dimensional analysis (see \cite{pozrikidis1992boundary}), the value of $\Sigma$ is defined by the geometry of the cell and the shear stress of the applied flow: $\Sigma\sim-\dot{\gamma}\eta_0R^3.$ Here $\eta_0$ is the viscosity of the suspending fluid.
Far from the cell, the flow in the midplane has the structure of  a 2D potential flow (Hele-Shaw flow) \cite{Liron1976}. Hence, a force would create a flow equivalent to  a 2D source dipole. Due to the superposition principle, a  force dipole  creates a flow
corresponding to a 2D source quadrupole, described by:
\begin{equation}\label{HSy}\mathbf{u}^{(Q)}\sim\frac{W\Sigma[Z(Z^2-3Y^2)\mathbf{\hat Z}-Y(Y^2-3Z^2)\mathbf{\hat Y}]}{\eta_0(Y^2+Z^2)^3}\left(1-\frac{4X^2}{W^2}\right),\end{equation} where $\mathbf{u}^{(Q)}$ is the flow velocity, $\mathbf{\hat Y}$ and $\mathbf{\hat Z}$ are the respective unit vectors and superscript $Q$ stands for quadrupolar. For $Y^2+Z^2>W^2,$ this equation well approximates the flow field of the numerical calculations in the $Y$-$Z$-plane (Fig.\,2B of the main text).
As shown in Fig.\,2B of the main text, the second cell is attracted to a position $Y = 0$. To assess the establishment of the equilibrium in $X$ and $Z$ directions, we use Eq. \ref{HSy} assuming $Y = 0$.

The velocity $\mathbf{u}$ of a second cell placed at $(0,0,\Delta Z)$ reads
$u_Z^{(Q)}\sim\Sigma W/(\eta_0 \Delta Z^3).$
Because $\Sigma$ is negative, this equation explains cell attraction, but a short-range repulsive interaction is also needed for a stable configuration to exist.
The answer arises from a subtle combination of the hydrodynamic interaction (i) of the cell at positive $\Delta Z$ with the cell at the origin and (ii) of both cells with the confining walls.

If well separated, cells stay in the midplane. However, as they approach each other due to the quadrupolar flow field, they are slightly pushed out of the midplane by short-range hydrodynamic interactions. By the symmetry of the problem, the displacement of the two cells from the midplane should be opposite. We therefore assume the first cell to be located at position $(-\Delta X,0,0)$ and the second cell to be located at position $(\Delta X,0,\Delta Z).$ For $\Delta X\neq 0,$ the $Z$-component of the velocity of the second cell has a contribution from the undisturbed shear flow $\dot\gamma \Delta X.$ For $\Delta X>0,$ this second contribution can balance the first one, as is shown below.

The $X$ component of the cell velocity has two contributions: (i) hydrodynamic interaction with the other cell and (ii) wall-induced migration. Hydrodynamic interactions of particles in a fluid domain bounded by two parallel walls have been treated theoretically \cite{Liron1976}. From the complicated formulae, we extracted the leading effect along the $X$ direction in a simple exploitable form.
Our calculations show that the $X$-component of the velocity induced by the first cell at the position of the second cell reads for $\Delta Z>0:$  \begin{equation}\label{ux}u_X^{(S)} \sim -\Sigma\sqrt{\Delta Z/W} e^{-b\Delta Z/W}\cos( a\Delta Z/W-\phi)/(\eta_0 W^2),\end{equation}  where superscript $S$ stands for short-range, $a\simeq 2.25$, $b\simeq 4.21$ and $\phi\simeq 2.49.$
$u_X^{(S)}$ vanishes at an infinite sequence of values of $\Delta Z.$
The velocity of the wall-induced migration scales as $u_X^{(L)}\sim \Sigma\Delta X/W^3$ \cite{farutin2013}, where the superscript $L$ stands for lift.

An equilibrium state will be attained when both the $Z$ and $X$ components of the test cell velocity are zero: $u_Z^{(Q)}+\dot{\gamma}\Delta X_0=0$ and $u_X^{(S)}+ u_X^{(L)}=0.$ {The former equation is represented by the red curve in Fig.\,4 of the main text, while the latter corresponds to the blue curve. Solving this system yields}
${C\Sigma W}/(\eta_0 \Delta Z_0^3)+\dot{\gamma}\Delta X_0=0$ and $\Delta X_0=C^\prime W^{3/2} \Delta Z_0^{-1/2} e^{-b\Delta Z_0/W}\cos( a\Delta Z_0/W-\phi),$ where $C$ and $C^\prime$ are numerical prefactors of order 1. A simple manipulation of these two equations yields the following final result
\begin{equation}\label{scaling}(R/W)^{3}=C^{\prime\prime}(\Delta Z_0/W)^{5/2}e^{-b\Delta Z_0/W}\cos(a \Delta Z_0/W-\phi), \end{equation} where $C^{\prime\prime}$ is another numerical prefactor, and we have used the scaling $\Sigma\sim-\eta_0\dot{\gamma}R^3.$ In Fig.\,4 of the main text, the two solutions of this equation are represented by a magenta circle and a purple square.
For a large enough width $W$ (as compared to $R$)  the left hand side becomes small compared to 1, so that
the solutions of equation \ref{scaling} are well approximated by zeros of the cosine function. This reasoning gives the scaling \begin{equation}\label{result}\Delta Z_0/W=\frac{\pi/2+\phi}{a}=1.805\end{equation} for the stable intercellular distance, independent of details (for example there is no need to specify the values of the prefactors $C$ and $C'$). The scaling (\ref{result}) defines
the position of the point $B$ in Fig.\,4 of the main text. The point $A$ in Fig.\,4 of the main text, corresponding to scaling $\Delta Z_A/W=0.409,$ proves to define an unstable stationary separation, as shown in Fig.\,4A of the main text. During approach, the second cell essentially follows the red curve until finally spiraling into point $B$ as shown in Fig.\,4B of the main text.

Reporting (\ref{scaling}) into the $X$-component of the equilibrium condition, yields $\Delta X_0/W = \tilde C  (R/W)^3>0,$ where $\tilde C$ is a positive numerical prefactor. The sign is such that the cell with $Z=\Delta Z_0$ is slightly above the midplane, while the one with $Z=0$ is slightly below. This result is in agreement with the full numerical simulation.

\begin{acknowledgments}
We are very grateful to V. Marchenko for many stimulating discussions. We thank Q. Xie, M. Wouters and D. Hessling for the help in numerics. C.M., A.F. and Z.S. thank the CNES (Centre National d'Etudes Spatiales) for a partial financial support and the French-German University Programme "Living Fluids" (Grant CFDA-Q1-14).  T.F. thanks the LIPhy laboratory, where the experimental work was performed, for financial support. The CPU time was provided by the HLRS (High-Performance Computing Center Stuttgart).
\end{acknowledgments}

\bibliography{ref}

\begin{thebibliography}{47}%
\makeatletter
\providecommand \@ifxundefined [1]{%
 \@ifx{#1\undefined}
}%
\providecommand \@ifnum [1]{%
 \ifnum #1\expandafter \@firstoftwo
 \else \expandafter \@secondoftwo
 \fi
}%
\providecommand \@ifx [1]{%
 \ifx #1\expandafter \@firstoftwo
 \else \expandafter \@secondoftwo
 \fi
}%
\providecommand \natexlab [1]{#1}%
\providecommand \enquote  [1]{``#1''}%
\providecommand \bibnamefont  [1]{#1}%
\providecommand \bibfnamefont [1]{#1}%
\providecommand \citenamefont [1]{#1}%
\providecommand \href@noop [0]{\@secondoftwo}%
\providecommand \href [0]{\begingroup \@sanitize@url \@href}%
\providecommand \@href[1]{\@@startlink{#1}\@@href}%
\providecommand \@@href[1]{\endgroup#1\@@endlink}%
\providecommand \@sanitize@url [0]{\catcode `\\12\catcode `\$12\catcode
  `\&12\catcode `\#12\catcode `\^12\catcode `\_12\catcode `\%12\relax}%
\providecommand \@@startlink[1]{}%
\providecommand \@@endlink[0]{}%
\providecommand \url  [0]{\begingroup\@sanitize@url \@url }%
\providecommand \@url [1]{\endgroup\@href {#1}{\urlprefix }}%
\providecommand \urlprefix  [0]{URL }%
\providecommand \Eprint [0]{\href }%
\providecommand \doibase [0]{http://dx.doi.org/}%
\providecommand \selectlanguage [0]{\@gobble}%
\providecommand \bibinfo  [0]{\@secondoftwo}%
\providecommand \bibfield  [0]{\@secondoftwo}%
\providecommand \translation [1]{[#1]}%
\providecommand \BibitemOpen [0]{}%
\providecommand \bibitemStop [0]{}%
\providecommand \bibitemNoStop [0]{.\EOS\space}%
\providecommand \EOS [0]{\spacefactor3000\relax}%
\providecommand \BibitemShut  [1]{\csname bibitem#1\endcsname}%
\let\auto@bib@innerbib\@empty
\bibitem [{\citenamefont {Popel}\ and\ \citenamefont
  {Johnson}(2005)}]{Popel:2005}%
  \BibitemOpen
  \bibfield  {author} {\bibinfo {author} {\bibfnamefont {A.~S.}\ \bibnamefont
  {Popel}}\ and\ \bibinfo {author} {\bibfnamefont {P.~C.}\ \bibnamefont
  {Johnson}},\ }\href@noop {} {\bibfield  {journal} {\bibinfo  {journal} {Annu
  Rev Fluid Mech}\ }\textbf {\bibinfo {volume} {37}},\ \bibinfo {pages} {43}
  (\bibinfo {year} {2005})}\BibitemShut {NoStop}%
\bibitem [{\citenamefont {Suresh}(2006)}]{Suresh2006}%
  \BibitemOpen
  \bibfield  {author} {\bibinfo {author} {\bibfnamefont {S.}~\bibnamefont
  {Suresh}},\ }\href@noop {} {\bibfield  {journal} {\bibinfo  {journal} {J
  Mater Res}\ }\textbf {\bibinfo {volume} {21}},\ \bibinfo {pages} {1871}
  (\bibinfo {year} {2006})}\BibitemShut {NoStop}%
\bibitem [{\citenamefont {Abkarian}\ \emph {et~al.}(2008)\citenamefont
  {Abkarian}, \citenamefont {Faivre}, \citenamefont {Horton}, \citenamefont
  {Smistrup}, \citenamefont {Best-Popescu},\ and\ \citenamefont
  {Stone}}]{AbkarianBMrev:2008}%
  \BibitemOpen
  \bibfield  {author} {\bibinfo {author} {\bibfnamefont {M.}~\bibnamefont
  {Abkarian}}, \bibinfo {author} {\bibfnamefont {M.}~\bibnamefont {Faivre}},
  \bibinfo {author} {\bibfnamefont {R.}~\bibnamefont {Horton}}, \bibinfo
  {author} {\bibfnamefont {K.}~\bibnamefont {Smistrup}}, \bibinfo {author}
  {\bibfnamefont {C.~A.}\ \bibnamefont {Best-Popescu}}, \ and\ \bibinfo
  {author} {\bibfnamefont {H.~A.}\ \bibnamefont {Stone}},\ }\href@noop {}
  {\bibfield  {journal} {\bibinfo  {journal} {Biomed Mater}\ }\textbf {\bibinfo
  {volume} {3}},\ \bibinfo {pages} {034011} (\bibinfo {year}
  {2008})}\BibitemShut {NoStop}%
\bibitem [{\citenamefont {Abkarian}\ and\ \citenamefont
  {Viallat}(2008)}]{Abkarian:2008}%
  \BibitemOpen
  \bibfield  {author} {\bibinfo {author} {\bibfnamefont {M.}~\bibnamefont
  {Abkarian}}\ and\ \bibinfo {author} {\bibfnamefont {A.}~\bibnamefont
  {Viallat}},\ }\href@noop {} {\bibfield  {journal} {\bibinfo  {journal} {Soft
  Matter}\ }\textbf {\bibinfo {volume} {4}},\ \bibinfo {pages} {653} (\bibinfo
  {year} {2008})}\BibitemShut {NoStop}%
\bibitem [{\citenamefont {Vlahovska}\ \emph
  {et~al.}(2009{\natexlab{a}})\citenamefont {Vlahovska}, \citenamefont
  {Podgorski},\ and\ \citenamefont {Misbah}}]{VlahovskaCR}%
  \BibitemOpen
  \bibfield  {author} {\bibinfo {author} {\bibfnamefont {P.~M.}\ \bibnamefont
  {Vlahovska}}, \bibinfo {author} {\bibfnamefont {T.}~\bibnamefont
  {Podgorski}}, \ and\ \bibinfo {author} {\bibfnamefont {C.}~\bibnamefont
  {Misbah}},\ }\href@noop {} {\bibfield  {journal} {\bibinfo  {journal} {C R
  Physique}\ }\textbf {\bibinfo {volume} {10}},\ \bibinfo {pages} {775}
  (\bibinfo {year} {2009}{\natexlab{a}})}\BibitemShut {NoStop}%
\bibitem [{\citenamefont {Fedosov}\ \emph {et~al.}(2011)\citenamefont
  {Fedosov}, \citenamefont {Pan}, \citenamefont {Caswell}, \citenamefont
  {Gompper},\ and\ \citenamefont {Karniadakis}}]{Fedosov:2011}%
  \BibitemOpen
  \bibfield  {author} {\bibinfo {author} {\bibfnamefont {D.~A.}\ \bibnamefont
  {Fedosov}}, \bibinfo {author} {\bibfnamefont {W.~X.}\ \bibnamefont {Pan}},
  \bibinfo {author} {\bibfnamefont {B.}~\bibnamefont {Caswell}}, \bibinfo
  {author} {\bibfnamefont {G.}~\bibnamefont {Gompper}}, \ and\ \bibinfo
  {author} {\bibfnamefont {G.~E.}\ \bibnamefont {Karniadakis}},\ }\href@noop {}
  {\bibfield  {journal} {\bibinfo  {journal} {Proc Nat Acad Sci}\ }\textbf
  {\bibinfo {volume} {108}},\ \bibinfo {pages} {11772} (\bibinfo {year}
  {2011})}\BibitemShut {NoStop}%
\bibitem [{\citenamefont {Li}\ \emph {et~al.}(2012)\citenamefont {Li},
  \citenamefont {Vlahovska},\ and\ \citenamefont
  {Karniadakis}}]{Li-Vlahovska:2012}%
  \BibitemOpen
  \bibfield  {author} {\bibinfo {author} {\bibfnamefont {X.}~\bibnamefont
  {Li}}, \bibinfo {author} {\bibfnamefont {P.~M.}\ \bibnamefont {Vlahovska}}, \
  and\ \bibinfo {author} {\bibfnamefont {G.~E.}\ \bibnamefont {Karniadakis}},\
  }\href@noop {} {\bibfield  {journal} {\bibinfo  {journal} {Soft Matter}\
  }\textbf {\bibinfo {volume} {9}},\ \bibinfo {pages} {28} (\bibinfo {year}
  {2012})}\BibitemShut {NoStop}%
\bibitem [{\citenamefont {Dupire}\ \emph {et~al.}(2012)\citenamefont {Dupire},
  \citenamefont {Socol},\ and\ \citenamefont {Viallat}}]{Dupire:2012}%
  \BibitemOpen
  \bibfield  {author} {\bibinfo {author} {\bibfnamefont {J.}~\bibnamefont
  {Dupire}}, \bibinfo {author} {\bibfnamefont {M.}~\bibnamefont {Socol}}, \
  and\ \bibinfo {author} {\bibfnamefont {A.}~\bibnamefont {Viallat}},\ }\href
  {\doibase 10.1073/pnas.1210236109} {\bibfield  {journal} {\bibinfo  {journal}
  {Proc Nat Acad Sci}\ }\textbf {\bibinfo {volume} {109}},\ \bibinfo {pages}
  {20808} (\bibinfo {year} {2012})}\BibitemShut {NoStop}%
\bibitem [{\citenamefont {Lanotte}\ \emph {et~al.}(2016)\citenamefont
  {Lanotte}, \citenamefont {Mauer}, \citenamefont {Mendez}, \citenamefont
  {Fedosov}, \citenamefont {Fromental}, \citenamefont {Claveria}, \citenamefont
  {Nicoud}, \citenamefont {Gompper},\ and\ \citenamefont
  {Abkarian}}]{Lanotte:2016}%
  \BibitemOpen
  \bibfield  {author} {\bibinfo {author} {\bibfnamefont {L.}~\bibnamefont
  {Lanotte}}, \bibinfo {author} {\bibfnamefont {J.}~\bibnamefont {Mauer}},
  \bibinfo {author} {\bibfnamefont {S.}~\bibnamefont {Mendez}}, \bibinfo
  {author} {\bibfnamefont {D.~A.}\ \bibnamefont {Fedosov}}, \bibinfo {author}
  {\bibfnamefont {J.-M.}\ \bibnamefont {Fromental}}, \bibinfo {author}
  {\bibfnamefont {V.}~\bibnamefont {Claveria}}, \bibinfo {author}
  {\bibfnamefont {F.}~\bibnamefont {Nicoud}}, \bibinfo {author} {\bibfnamefont
  {G.}~\bibnamefont {Gompper}}, \ and\ \bibinfo {author} {\bibfnamefont
  {M.}~\bibnamefont {Abkarian}},\ }\href {\doibase 10.1073/pnas.1608074113}
  {\bibfield  {journal} {\bibinfo  {journal} {Proc Nat Acad Sci}\ }\textbf
  {\bibinfo {volume} {113}},\ \bibinfo {pages} {13289} (\bibinfo {year}
  {2016})}\BibitemShut {NoStop}%
\bibitem [{\citenamefont {McWhirter}\ \emph
  {et~al.}(2009{\natexlab{a}})\citenamefont {McWhirter}, \citenamefont
  {Noguchi},\ and\ \citenamefont {Gompper}}]{McWhirter-Hiroshi-Gompper:2009}%
  \BibitemOpen
  \bibfield  {author} {\bibinfo {author} {\bibfnamefont {J.~L.}\ \bibnamefont
  {McWhirter}}, \bibinfo {author} {\bibfnamefont {H.}~\bibnamefont {Noguchi}},
  \ and\ \bibinfo {author} {\bibfnamefont {G.}~\bibnamefont {Gompper}},\ }\href
  {\doibase 10.1073/pnas.0811484106} {\bibfield  {journal} {\bibinfo  {journal}
  {Proc Nat Acad Sci}\ }\textbf {\bibinfo {volume} {106}},\ \bibinfo {pages}
  {6039} (\bibinfo {year} {2009}{\natexlab{a}})}\BibitemShut {NoStop}%
\bibitem [{\citenamefont {McWhirter}\ \emph {et~al.}(2012)\citenamefont
  {McWhirter}, \citenamefont {Noguchi},\ and\ \citenamefont
  {Gompper}}]{McWhirter:2012}%
  \BibitemOpen
  \bibfield  {author} {\bibinfo {author} {\bibfnamefont {J.~L.}\ \bibnamefont
  {McWhirter}}, \bibinfo {author} {\bibfnamefont {H.}~\bibnamefont {Noguchi}},
  \ and\ \bibinfo {author} {\bibfnamefont {G.}~\bibnamefont {Gompper}},\
  }\href@noop {} {\bibfield  {journal} {\bibinfo  {journal} {New J Phys}\
  }\textbf {\bibinfo {volume} {14}},\ \bibinfo {pages} {085026} (\bibinfo
  {year} {2012})}\BibitemShut {NoStop}%
\bibitem [{\citenamefont {Tomaiuolo}\ \emph {et~al.}(2012)\citenamefont
  {Tomaiuolo}, \citenamefont {Lanotte}, \citenamefont {Ghigliotti},
  \citenamefont {Misbah},\ and\ \citenamefont {Guido}}]{Tomaiuolo:2012}%
  \BibitemOpen
  \bibfield  {author} {\bibinfo {author} {\bibfnamefont {G.}~\bibnamefont
  {Tomaiuolo}}, \bibinfo {author} {\bibfnamefont {L.}~\bibnamefont {Lanotte}},
  \bibinfo {author} {\bibfnamefont {G.}~\bibnamefont {Ghigliotti}}, \bibinfo
  {author} {\bibfnamefont {C.}~\bibnamefont {Misbah}}, \ and\ \bibinfo {author}
  {\bibfnamefont {S.}~\bibnamefont {Guido}},\ }\href@noop {} {\bibfield
  {journal} {\bibinfo  {journal} {Phys Fluids}\ }\textbf {\bibinfo {volume}
  {24}},\ \bibinfo {pages} {051903} (\bibinfo {year} {2012})}\BibitemShut
  {NoStop}%
\bibitem [{\citenamefont {Beatus}\ \emph {et~al.}(2006)\citenamefont {Beatus},
  \citenamefont {Tlusty},\ and\ \citenamefont {Bar-Ziv}}]{beatus2006phonons}%
  \BibitemOpen
  \bibfield  {author} {\bibinfo {author} {\bibfnamefont {T.}~\bibnamefont
  {Beatus}}, \bibinfo {author} {\bibfnamefont {T.}~\bibnamefont {Tlusty}}, \
  and\ \bibinfo {author} {\bibfnamefont {R.}~\bibnamefont {Bar-Ziv}},\
  }\href@noop {} {\bibfield  {journal} {\bibinfo  {journal} {Nat Phys}\
  }\textbf {\bibinfo {volume} {2}},\ \bibinfo {pages} {743} (\bibinfo {year}
  {2006})}\BibitemShut {NoStop}%
\bibitem [{\citenamefont {Beatus}\ \emph {et~al.}(2007)\citenamefont {Beatus},
  \citenamefont {Bar-Ziv},\ and\ \citenamefont {Tlusty}}]{beatus2007anomalous}%
  \BibitemOpen
  \bibfield  {author} {\bibinfo {author} {\bibfnamefont {T.}~\bibnamefont
  {Beatus}}, \bibinfo {author} {\bibfnamefont {R.}~\bibnamefont {Bar-Ziv}}, \
  and\ \bibinfo {author} {\bibfnamefont {T.}~\bibnamefont {Tlusty}},\
  }\href@noop {} {\bibfield  {journal} {\bibinfo  {journal} {Phys Rev Lett}\
  }\textbf {\bibinfo {volume} {99}},\ \bibinfo {pages} {124502} (\bibinfo
  {year} {2007})}\BibitemShut {NoStop}%
\bibitem [{\citenamefont {Beatus}\ \emph {et~al.}(2012)\citenamefont {Beatus},
  \citenamefont {Bar-Ziv},\ and\ \citenamefont {Tlusty}}]{beatus2012physics}%
  \BibitemOpen
  \bibfield  {author} {\bibinfo {author} {\bibfnamefont {T.}~\bibnamefont
  {Beatus}}, \bibinfo {author} {\bibfnamefont {R.~H.}\ \bibnamefont {Bar-Ziv}},
  \ and\ \bibinfo {author} {\bibfnamefont {T.}~\bibnamefont {Tlusty}},\
  }\href@noop {} {\bibfield  {journal} {\bibinfo  {journal} {Phys Rep}\
  }\textbf {\bibinfo {volume} {516}},\ \bibinfo {pages} {103} (\bibinfo {year}
  {2012})}\BibitemShut {NoStop}%
\bibitem [{\citenamefont {Uspal}\ and\ \citenamefont
  {Doyle}(2012)}]{uspal2012collective}%
  \BibitemOpen
  \bibfield  {author} {\bibinfo {author} {\bibfnamefont {W.~E.}\ \bibnamefont
  {Uspal}}\ and\ \bibinfo {author} {\bibfnamefont {P.~S.}\ \bibnamefont
  {Doyle}},\ }\href@noop {} {\bibfield  {journal} {\bibinfo  {journal} {Soft
  Matter}\ }\textbf {\bibinfo {volume} {8}},\ \bibinfo {pages} {10676}
  (\bibinfo {year} {2012})}\BibitemShut {NoStop}%
\bibitem [{\citenamefont {Janssen}\ \emph {et~al.}(2012)\citenamefont
  {Janssen}, \citenamefont {Baron}, \citenamefont {Anderson}, \citenamefont
  {Blawzdziewicz}, \citenamefont {Loewenberg},\ and\ \citenamefont
  {Wajnryb}}]{janssen2012collective}%
  \BibitemOpen
  \bibfield  {author} {\bibinfo {author} {\bibfnamefont {P.}~\bibnamefont
  {Janssen}}, \bibinfo {author} {\bibfnamefont {M.}~\bibnamefont {Baron}},
  \bibinfo {author} {\bibfnamefont {P.}~\bibnamefont {Anderson}}, \bibinfo
  {author} {\bibfnamefont {J.}~\bibnamefont {Blawzdziewicz}}, \bibinfo {author}
  {\bibfnamefont {M.}~\bibnamefont {Loewenberg}}, \ and\ \bibinfo {author}
  {\bibfnamefont {E.}~\bibnamefont {Wajnryb}},\ }\href@noop {} {\bibfield
  {journal} {\bibinfo  {journal} {Soft Matter}\ }\textbf {\bibinfo {volume}
  {8}},\ \bibinfo {pages} {7495} (\bibinfo {year} {2012})}\BibitemShut
  {NoStop}%
\bibitem [{\citenamefont {Shani}\ \emph
  {et~al.}(2014{\natexlab{a}})\citenamefont {Shani}, \citenamefont {Beatus},
  \citenamefont {Bar-Ziv},\ and\ \citenamefont {Tlusty}}]{shani2014long}%
  \BibitemOpen
  \bibfield  {author} {\bibinfo {author} {\bibfnamefont {I.}~\bibnamefont
  {Shani}}, \bibinfo {author} {\bibfnamefont {T.}~\bibnamefont {Beatus}},
  \bibinfo {author} {\bibfnamefont {R.~H.}\ \bibnamefont {Bar-Ziv}}, \ and\
  \bibinfo {author} {\bibfnamefont {T.}~\bibnamefont {Tlusty}},\ }\href@noop {}
  {\bibfield  {journal} {\bibinfo  {journal} {Nat Phys}\ }\textbf {\bibinfo
  {volume} {10}},\ \bibinfo {pages} {140} (\bibinfo {year}
  {2014}{\natexlab{a}})}\BibitemShut {NoStop}%
\bibitem [{\citenamefont {Uspal}\ and\ \citenamefont
  {Doyle}(2014)}]{uspal2014self}%
  \BibitemOpen
  \bibfield  {author} {\bibinfo {author} {\bibfnamefont {W.~E.}\ \bibnamefont
  {Uspal}}\ and\ \bibinfo {author} {\bibfnamefont {P.~S.}\ \bibnamefont
  {Doyle}},\ }\href@noop {} {\bibfield  {journal} {\bibinfo  {journal} {Soft
  matter}\ }\textbf {\bibinfo {volume} {10}},\ \bibinfo {pages} {5177}
  (\bibinfo {year} {2014})}\BibitemShut {NoStop}%
\bibitem [{\citenamefont {da~Cunha}\ and\ \citenamefont
  {Hinch}(1996)}]{Cunha1996}%
  \BibitemOpen
  \bibfield  {author} {\bibinfo {author} {\bibfnamefont {F.}~\bibnamefont
  {da~Cunha}}\ and\ \bibinfo {author} {\bibfnamefont {E.}~\bibnamefont
  {Hinch}},\ }\href@noop {} {\bibfield  {journal} {\bibinfo  {journal} {J Fluid
  Mech}\ }\textbf {\bibinfo {volume} {309}},\ \bibinfo {pages} {211} (\bibinfo
  {year} {1996})}\BibitemShut {NoStop}%
\bibitem [{\citenamefont {Zhao}\ \emph {et~al.}(2012)\citenamefont {Zhao},
  \citenamefont {Shaqfeh},\ and\ \citenamefont {Narsimhan}}]{zhao2012}%
  \BibitemOpen
  \bibfield  {author} {\bibinfo {author} {\bibfnamefont {H.}~\bibnamefont
  {Zhao}}, \bibinfo {author} {\bibfnamefont {E.~S.}\ \bibnamefont {Shaqfeh}}, \
  and\ \bibinfo {author} {\bibfnamefont {V.}~\bibnamefont {Narsimhan}},\
  }\href@noop {} {\bibfield  {journal} {\bibinfo  {journal} {Phys Fluids}\
  }\textbf {\bibinfo {volume} {24}},\ \bibinfo {pages} {011902} (\bibinfo
  {year} {2012})}\BibitemShut {NoStop}%
\bibitem [{\citenamefont {Grandchamp}\ \emph {et~al.}(2013)\citenamefont
  {Grandchamp}, \citenamefont {Coupier}, \citenamefont {Srivastav},
  \citenamefont {Minetti},\ and\ \citenamefont {Podgorski}}]{grandchamp2013}%
  \BibitemOpen
  \bibfield  {author} {\bibinfo {author} {\bibfnamefont {X.}~\bibnamefont
  {Grandchamp}}, \bibinfo {author} {\bibfnamefont {G.}~\bibnamefont {Coupier}},
  \bibinfo {author} {\bibfnamefont {A.}~\bibnamefont {Srivastav}}, \bibinfo
  {author} {\bibfnamefont {C.}~\bibnamefont {Minetti}}, \ and\ \bibinfo
  {author} {\bibfnamefont {T.}~\bibnamefont {Podgorski}},\ }\href@noop {}
  {\bibfield  {journal} {\bibinfo  {journal} {Phys Rev Lett}\ }\textbf
  {\bibinfo {volume} {110}},\ \bibinfo {pages} {108101} (\bibinfo {year}
  {2013})}\BibitemShut {NoStop}%
\bibitem [{\citenamefont {Henr\'{\i}quez~Rivera}\ \emph
  {et~al.}(2015)\citenamefont {Henr\'{\i}quez~Rivera}, \citenamefont {Sinha},\
  and\ \citenamefont {Graham}}]{rivera2015}%
  \BibitemOpen
  \bibfield  {author} {\bibinfo {author} {\bibfnamefont {R.~G.}\ \bibnamefont
  {Henr\'{\i}quez~Rivera}}, \bibinfo {author} {\bibfnamefont {K.}~\bibnamefont
  {Sinha}}, \ and\ \bibinfo {author} {\bibfnamefont {M.~D.}\ \bibnamefont
  {Graham}},\ }\href {\doibase 10.1103/PhysRevLett.114.188101} {\bibfield
  {journal} {\bibinfo  {journal} {Phys Rev Lett}\ }\textbf {\bibinfo {volume}
  {114}},\ \bibinfo {pages} {188101} (\bibinfo {year} {2015})}\BibitemShut
  {NoStop}%
\bibitem [{\citenamefont {Baron}\ \emph {et~al.}(2008)\citenamefont {Baron},
  \citenamefont {B{\l}awzdziewicz},\ and\ \citenamefont
  {Wajnryb}}]{baron2008hydrodynamic}%
  \BibitemOpen
  \bibfield  {author} {\bibinfo {author} {\bibfnamefont {M.}~\bibnamefont
  {Baron}}, \bibinfo {author} {\bibfnamefont {J.}~\bibnamefont
  {B{\l}awzdziewicz}}, \ and\ \bibinfo {author} {\bibfnamefont
  {E.}~\bibnamefont {Wajnryb}},\ }\href@noop {} {\bibfield  {journal} {\bibinfo
   {journal} {Phys Rev Lett}\ }\textbf {\bibinfo {volume} {100}},\ \bibinfo
  {pages} {174502} (\bibinfo {year} {2008})}\BibitemShut {NoStop}%
\bibitem [{\citenamefont {Lee}\ \emph {et~al.}(2010)\citenamefont {Lee},
  \citenamefont {Amini}, \citenamefont {Stone},\ and\ \citenamefont
  {Di~Carlo}}]{lee2010dynamic}%
  \BibitemOpen
  \bibfield  {author} {\bibinfo {author} {\bibfnamefont {W.}~\bibnamefont
  {Lee}}, \bibinfo {author} {\bibfnamefont {H.}~\bibnamefont {Amini}}, \bibinfo
  {author} {\bibfnamefont {H.~A.}\ \bibnamefont {Stone}}, \ and\ \bibinfo
  {author} {\bibfnamefont {D.}~\bibnamefont {Di~Carlo}},\ }\href@noop {}
  {\bibfield  {journal} {\bibinfo  {journal} {Proc Nat Acad Sci}\ }\textbf
  {\bibinfo {volume} {107}},\ \bibinfo {pages} {22413} (\bibinfo {year}
  {2010})}\BibitemShut {NoStop}%
\bibitem [{\citenamefont {Humphry}\ \emph {et~al.}(2010)\citenamefont
  {Humphry}, \citenamefont {Kulkarni}, \citenamefont {Weitz}, \citenamefont
  {Morris},\ and\ \citenamefont {Stone}}]{humphry2010axial}%
  \BibitemOpen
  \bibfield  {author} {\bibinfo {author} {\bibfnamefont {K.~J.}\ \bibnamefont
  {Humphry}}, \bibinfo {author} {\bibfnamefont {P.~M.}\ \bibnamefont
  {Kulkarni}}, \bibinfo {author} {\bibfnamefont {D.~A.}\ \bibnamefont {Weitz}},
  \bibinfo {author} {\bibfnamefont {J.~F.}\ \bibnamefont {Morris}}, \ and\
  \bibinfo {author} {\bibfnamefont {H.~A.}\ \bibnamefont {Stone}},\ }\href@noop
  {} {\bibfield  {journal} {\bibinfo  {journal} {Phys Fluids}\ }\textbf
  {\bibinfo {volume} {22}},\ \bibinfo {pages} {081703} (\bibinfo {year}
  {2010})}\BibitemShut {NoStop}%
\bibitem [{\citenamefont {Shani}\ \emph
  {et~al.}(2014{\natexlab{b}})\citenamefont {Shani}, \citenamefont {Beatus},
  \citenamefont {Bar-Ziv},\ and\ \citenamefont {Tlusty}}]{Shani:2014}%
  \BibitemOpen
  \bibfield  {author} {\bibinfo {author} {\bibfnamefont {I.}~\bibnamefont
  {Shani}}, \bibinfo {author} {\bibfnamefont {T.}~\bibnamefont {Beatus}},
  \bibinfo {author} {\bibfnamefont {R.~H.}\ \bibnamefont {Bar-Ziv}}, \ and\
  \bibinfo {author} {\bibfnamefont {T.}~\bibnamefont {Tlusty}},\ }\href
  {\doibase 10.1038/NPHYS2843} {\bibfield  {journal} {\bibinfo  {journal} {Nat
  Phys}\ }\textbf {\bibinfo {volume} {10}},\ \bibinfo {pages} {140} (\bibinfo
  {year} {2014}{\natexlab{b}})}\BibitemShut {NoStop}%
\bibitem [{\citenamefont {Bricard}\ \emph {et~al.}(2013)\citenamefont
  {Bricard}, \citenamefont {Caussin}, \citenamefont {Desreumaux}, \citenamefont
  {Dauchot},\ and\ \citenamefont {Bartolo}}]{Bartolo:2013}%
  \BibitemOpen
  \bibfield  {author} {\bibinfo {author} {\bibfnamefont {A.}~\bibnamefont
  {Bricard}}, \bibinfo {author} {\bibfnamefont {J.-B.}\ \bibnamefont
  {Caussin}}, \bibinfo {author} {\bibfnamefont {N.}~\bibnamefont {Desreumaux}},
  \bibinfo {author} {\bibfnamefont {O.}~\bibnamefont {Dauchot}}, \ and\
  \bibinfo {author} {\bibfnamefont {D.}~\bibnamefont {Bartolo}},\ }\href
  {\doibase 10.1038/nature12673} {\bibfield  {journal} {\bibinfo  {journal}
  {Nature}\ }\textbf {\bibinfo {volume} {503}},\ \bibinfo {pages} {95}
  (\bibinfo {year} {2013})}\BibitemShut {NoStop}%
\bibitem [{\citenamefont {Bricard}\ \emph {et~al.}(2015)\citenamefont
  {Bricard}, \citenamefont {Caussin}, \citenamefont {Das}, \citenamefont
  {Savoie}, \citenamefont {Chikkadi}, \citenamefont {Shitara}, \citenamefont
  {Chepizhko}, \citenamefont {Peruani}, \citenamefont {Saintillan},\ and\
  \citenamefont {Bartolo}}]{Bartolo:2015}%
  \BibitemOpen
  \bibfield  {author} {\bibinfo {author} {\bibfnamefont {A.}~\bibnamefont
  {Bricard}}, \bibinfo {author} {\bibfnamefont {J.-B.}\ \bibnamefont
  {Caussin}}, \bibinfo {author} {\bibfnamefont {D.}~\bibnamefont {Das}},
  \bibinfo {author} {\bibfnamefont {C.}~\bibnamefont {Savoie}}, \bibinfo
  {author} {\bibfnamefont {V.}~\bibnamefont {Chikkadi}}, \bibinfo {author}
  {\bibfnamefont {K.}~\bibnamefont {Shitara}}, \bibinfo {author} {\bibfnamefont
  {O.}~\bibnamefont {Chepizhko}}, \bibinfo {author} {\bibfnamefont
  {F.}~\bibnamefont {Peruani}}, \bibinfo {author} {\bibfnamefont
  {D.}~\bibnamefont {Saintillan}}, \ and\ \bibinfo {author} {\bibfnamefont
  {D.}~\bibnamefont {Bartolo}},\ }\href {\doibase 10.1038/ncomms8470}
  {\bibfield  {journal} {\bibinfo  {journal} {Nat Commun}\ }\textbf {\bibinfo
  {volume} {6}} (\bibinfo {year} {2015}),\ 10.1038/ncomms8470}\BibitemShut
  {NoStop}%
\bibitem [{\citenamefont {Yeo}\ \emph {et~al.}(2015)\citenamefont {Yeo},
  \citenamefont {Lushi},\ and\ \citenamefont {Vlahovska}}]{Yeo:2015}%
  \BibitemOpen
  \bibfield  {author} {\bibinfo {author} {\bibfnamefont {K.}~\bibnamefont
  {Yeo}}, \bibinfo {author} {\bibfnamefont {E.}~\bibnamefont {Lushi}}, \ and\
  \bibinfo {author} {\bibfnamefont {P.~M.}\ \bibnamefont {Vlahovska}},\
  }\href@noop {} {\bibfield  {journal} {\bibinfo  {journal} {Phys. Rev. Lett.}\
  }\textbf {\bibinfo {volume} {114}},\ \bibinfo {pages} {188301} (\bibinfo
  {year} {2015})}\BibitemShut {NoStop}%
\bibitem [{\citenamefont {Goto}\ and\ \citenamefont
  {Tanaka}(2015)}]{goto2015purely}%
  \BibitemOpen
  \bibfield  {author} {\bibinfo {author} {\bibfnamefont {Y.}~\bibnamefont
  {Goto}}\ and\ \bibinfo {author} {\bibfnamefont {H.}~\bibnamefont {Tanaka}},\
  }\href@noop {} {\bibfield  {journal} {\bibinfo  {journal} {Nat Commun}\
  }\textbf {\bibinfo {volume} {6}},\ \bibinfo {pages} {5994} (\bibinfo {year}
  {2015})}\BibitemShut {NoStop}%
\bibitem [{\citenamefont {Fischer}\ and\ \citenamefont
  {Richardson}(1980)}]{RBCtrain}%
  \BibitemOpen
  \bibfield  {author} {\bibinfo {author} {\bibfnamefont {T.}~\bibnamefont
  {Fischer}}\ and\ \bibinfo {author} {\bibfnamefont {P.}~\bibnamefont
  {Richardson}},\ }in\ \href@noop {} {\emph {\bibinfo {booktitle} {Amer. Soc
  Mech. Engrs. 1980 Advances in Bioengineering}}},\ \bibinfo {editor} {edited
  by\ \bibinfo {editor} {\bibfnamefont {V.}~\bibnamefont {Mow}}}\ (\bibinfo
  {year} {1980})\ pp.\ \bibinfo {pages} {305--308}\BibitemShut {NoStop}%
\bibitem [{\citenamefont {Bull}\ \emph {et~al.}(1983)\citenamefont {Bull},
  \citenamefont {Feo},\ and\ \citenamefont {Bessis}}]{Bull:1983}%
  \BibitemOpen
  \bibfield  {author} {\bibinfo {author} {\bibfnamefont {B.}~\bibnamefont
  {Bull}}, \bibinfo {author} {\bibfnamefont {C.}~\bibnamefont {Feo}}, \ and\
  \bibinfo {author} {\bibfnamefont {M.}~\bibnamefont {Bessis}},\ }\href@noop {}
  {\bibfield  {journal} {\bibinfo  {journal} {Cytometry}\ }\textbf {\bibinfo
  {volume} {3}},\ \bibinfo {pages} {300} (\bibinfo {year} {1983})}\BibitemShut
  {NoStop}%
\bibitem [{\citenamefont {Pathak}\ \emph {et~al.}(2002)\citenamefont {Pathak},
  \citenamefont {Davis}, \citenamefont {Hudson},\ and\ \citenamefont
  {Migler}}]{pathak2002layered}%
  \BibitemOpen
  \bibfield  {author} {\bibinfo {author} {\bibfnamefont {J.~A.}\ \bibnamefont
  {Pathak}}, \bibinfo {author} {\bibfnamefont {M.~C.}\ \bibnamefont {Davis}},
  \bibinfo {author} {\bibfnamefont {S.~D.}\ \bibnamefont {Hudson}}, \ and\
  \bibinfo {author} {\bibfnamefont {K.~B.}\ \bibnamefont {Migler}},\
  }\href@noop {} {\bibfield  {journal} {\bibinfo  {journal} {J Colloid
  Interface Sci}\ }\textbf {\bibinfo {volume} {255}},\ \bibinfo {pages} {391}
  (\bibinfo {year} {2002})}\BibitemShut {NoStop}%
\bibitem [{\citenamefont {Kr\"uger}\ \emph {et~al.}(2013)\citenamefont
  {Kr\"uger}, \citenamefont {Frijters}, \citenamefont {G\"unther},
  \citenamefont {Kaoui},\ and\ \citenamefont {Harting}}]{Krueger2013}%
  \BibitemOpen
  \bibfield  {author} {\bibinfo {author} {\bibfnamefont {T.}~\bibnamefont
  {Kr\"uger}}, \bibinfo {author} {\bibfnamefont {S.}~\bibnamefont {Frijters}},
  \bibinfo {author} {\bibfnamefont {F.}~\bibnamefont {G\"unther}}, \bibinfo
  {author} {\bibfnamefont {B.}~\bibnamefont {Kaoui}}, \ and\ \bibinfo {author}
  {\bibfnamefont {J.}~\bibnamefont {Harting}},\ }\href@noop {} {\bibfield
  {journal} {\bibinfo  {journal} {Eur Phys J Special Topics}\ }\textbf
  {\bibinfo {volume} {222}},\ \bibinfo {pages} {177} (\bibinfo {year}
  {2013})}\BibitemShut {NoStop}%
\bibitem [{\citenamefont {Kr{\"u}ger}\ \emph {et~al.}(2014)\citenamefont
  {Kr{\"u}ger}, \citenamefont {Kaoui},\ and\ \citenamefont
  {Harting}}]{kruger2014}%
  \BibitemOpen
  \bibfield  {author} {\bibinfo {author} {\bibfnamefont {T.}~\bibnamefont
  {Kr{\"u}ger}}, \bibinfo {author} {\bibfnamefont {B.}~\bibnamefont {Kaoui}}, \
  and\ \bibinfo {author} {\bibfnamefont {J.}~\bibnamefont {Harting}},\
  }\href@noop {} {\bibfield  {journal} {\bibinfo  {journal} {J Fluid Mech}\
  }\textbf {\bibinfo {volume} {751}},\ \bibinfo {pages} {725} (\bibinfo {year}
  {2014})}\BibitemShut {NoStop}%
\bibitem [{SM()}]{SM}%
  \BibitemOpen
  \href@noop {} {\bibinfo  {journal} {See Supplemental Material}\ }\BibitemShut
  {NoStop}%
\bibitem [{\citenamefont {Fischer}\ \emph {et~al.}(1978)\citenamefont
  {Fischer}, \citenamefont {St\"ohr-Liesen},\ and\ \citenamefont
  {Schmid-Sch\"onbein}}]{Fischer894}%
  \BibitemOpen
\bibfield  {journal} {  }\bibfield  {author} {\bibinfo {author} {\bibfnamefont
  {T.}~\bibnamefont {Fischer}}, \bibinfo {author} {\bibfnamefont
  {M.}~\bibnamefont {St\"ohr-Liesen}}, \ and\ \bibinfo {author} {\bibfnamefont
  {H.}~\bibnamefont {Schmid-Sch\"onbein}},\ }\href {\doibase
  10.1126/science.715448} {\bibfield  {journal} {\bibinfo  {journal} {Science}\
  }\textbf {\bibinfo {volume} {202}},\ \bibinfo {pages} {894} (\bibinfo {year}
  {1978})}\BibitemShut {NoStop}%
\bibitem [{\citenamefont {Cantat}\ and\ \citenamefont
  {Misbah}(1999)}]{Cantat1999b}%
  \BibitemOpen
  \bibfield  {author} {\bibinfo {author} {\bibfnamefont {I.}~\bibnamefont
  {Cantat}}\ and\ \bibinfo {author} {\bibfnamefont {C.}~\bibnamefont
  {Misbah}},\ }\href@noop {} {\bibfield  {journal} {\bibinfo  {journal} {Phys
  Rev Lett}\ }\textbf {\bibinfo {volume} {83}},\ \bibinfo {pages} {880}
  (\bibinfo {year} {1999})}\BibitemShut {NoStop}%
\bibitem [{\citenamefont {Seifert}(1999)}]{seifert1999}%
  \BibitemOpen
  \bibfield  {author} {\bibinfo {author} {\bibfnamefont {U.}~\bibnamefont
  {Seifert}},\ }\href@noop {} {\bibfield  {journal} {\bibinfo  {journal} {Phys
  Rev Lett}\ }\textbf {\bibinfo {volume} {83}},\ \bibinfo {pages} {876}
  (\bibinfo {year} {1999})}\BibitemShut {NoStop}%
\bibitem [{\citenamefont {Sukumaran}\ and\ \citenamefont
  {Seifert}(2001)}]{sukumaran2001}%
  \BibitemOpen
  \bibfield  {author} {\bibinfo {author} {\bibfnamefont {S.}~\bibnamefont
  {Sukumaran}}\ and\ \bibinfo {author} {\bibfnamefont {U.}~\bibnamefont
  {Seifert}},\ }\href@noop {} {\bibfield  {journal} {\bibinfo  {journal} {Phys
  Rev E}\ }\textbf {\bibinfo {volume} {64}},\ \bibinfo {pages} {011916}
  (\bibinfo {year} {2001})}\BibitemShut {NoStop}%
\bibitem [{\citenamefont {Zurita-Gotor}\ \emph {et~al.}(2007)\citenamefont
  {Zurita-Gotor}, \citenamefont {B{\l}awzdziewicz},\ and\ \citenamefont
  {Wajnryb}}]{zurita2007swapping}%
  \BibitemOpen
  \bibfield  {author} {\bibinfo {author} {\bibfnamefont {M.}~\bibnamefont
  {Zurita-Gotor}}, \bibinfo {author} {\bibfnamefont {J.}~\bibnamefont
  {B{\l}awzdziewicz}}, \ and\ \bibinfo {author} {\bibfnamefont
  {E.}~\bibnamefont {Wajnryb}},\ }\href@noop {} {\bibfield  {journal} {\bibinfo
   {journal} {J Fluid Mech}\ }\textbf {\bibinfo {volume} {592}},\ \bibinfo
  {pages} {447} (\bibinfo {year} {2007})}\BibitemShut {NoStop}%
\bibitem [{\citenamefont {McWhirter}\ \emph
  {et~al.}(2009{\natexlab{b}})\citenamefont {McWhirter}, \citenamefont
  {Noguchi},\ and\ \citenamefont {Gompper}}]{mcwhirter2009}%
  \BibitemOpen
  \bibfield  {author} {\bibinfo {author} {\bibfnamefont {J.~L.}\ \bibnamefont
  {McWhirter}}, \bibinfo {author} {\bibfnamefont {H.}~\bibnamefont {Noguchi}},
  \ and\ \bibinfo {author} {\bibfnamefont {G.}~\bibnamefont {Gompper}},\
  }\href@noop {} {\bibfield  {journal} {\bibinfo  {journal} {Proc Nat Acad
  Sci}\ }\textbf {\bibinfo {volume} {106}},\ \bibinfo {pages} {6039} (\bibinfo
  {year} {2009}{\natexlab{b}})}\BibitemShut {NoStop}%
\bibitem [{\citenamefont {Liron}\ and\ \citenamefont
  {Mochon}(1976)}]{Liron1976}%
  \BibitemOpen
  \bibfield  {author} {\bibinfo {author} {\bibfnamefont {N.}~\bibnamefont
  {Liron}}\ and\ \bibinfo {author} {\bibfnamefont {S.}~\bibnamefont {Mochon}},\
  }\href {\doibase 10.1007/BF01535565} {\bibfield  {journal} {\bibinfo
  {journal} {J Eng Math}\ }\textbf {\bibinfo {volume} {10}},\ \bibinfo {pages}
  {287} (\bibinfo {year} {1976})}\BibitemShut {NoStop}%
\bibitem [{\citenamefont {Vlahovska}\ \emph
  {et~al.}(2009{\natexlab{b}})\citenamefont {Vlahovska}, \citenamefont
  {Podgorski},\ and\ \citenamefont {Misbah}}]{Petia_review}%
  \BibitemOpen
  \bibfield  {author} {\bibinfo {author} {\bibfnamefont {P.~M.}\ \bibnamefont
  {Vlahovska}}, \bibinfo {author} {\bibfnamefont {T.}~\bibnamefont
  {Podgorski}}, \ and\ \bibinfo {author} {\bibfnamefont {C.}~\bibnamefont
  {Misbah}},\ }\href@noop {} {\bibfield  {journal} {\bibinfo  {journal} {C R
  Phys}\ }\textbf {\bibinfo {volume} {10}},\ \bibinfo {pages} {775} (\bibinfo
  {year} {2009}{\natexlab{b}})}\BibitemShut {NoStop}%
\bibitem [{\citenamefont {Pozrikidis}(1992)}]{pozrikidis1992boundary}%
  \BibitemOpen
  \bibfield  {author} {\bibinfo {author} {\bibfnamefont {C.}~\bibnamefont
  {Pozrikidis}},\ }\href@noop {} {\emph {\bibinfo {title} {Boundary integral
  and singularity methods for linearized viscous flow}}}\ (\bibinfo
  {publisher} {Cambridge University Press},\ \bibinfo {year}
  {1992})\BibitemShut {NoStop}%
\bibitem [{\citenamefont {Farutin}\ and\ \citenamefont
  {Misbah}(2013)}]{farutin2013}%
  \BibitemOpen
  \bibfield  {author} {\bibinfo {author} {\bibfnamefont {A.}~\bibnamefont
  {Farutin}}\ and\ \bibinfo {author} {\bibfnamefont {C.}~\bibnamefont
  {Misbah}},\ }\href@noop {} {\bibfield  {journal} {\bibinfo  {journal} {Phys
  Rev Lett}\ }\textbf {\bibinfo {volume} {110}},\ \bibinfo {pages} {108104}
  (\bibinfo {year} {2013})}\BibitemShut {NoStop}%
\end{thebibliography}%

\end{document}